\documentclass[10pt,prb,twocolumn,showpacs,superscriptaddress,floatfix]{revtex4}

\usepackage{graphicx}

\newcommand{\be}{\begin{equation}}
\newcommand{\ee}{\end{equation}}
\newcommand{\bea}{\begin{eqnarray}}
\newcommand{\eea}{\end{eqnarray}}

\newcommand{\f}{\frac}
\newcommand{\vdimer}{{\vrule height0.2cm width0.05cm depth0pt}}
\newcommand{\hdimer}{{\hrule height0.05cm width0.2cm depth0pt}}

\newcommand{\verdimers}{\hbox{\vdimer \hskip 0.1cm \vdimer}}
\newcommand{\hordimers}{\hbox{\vbox{\hdimer \vskip 0.1cm \hdimer}}}

\begin{document}

\title{The plaquette phase of the square lattice quantum dimer model}

\author{Olav  F. Sylju{\aa}sen}
\affiliation{NORDITA, Blegdamsvej 17, DK-2100 Copenhagen {\O}, Denmark}
\email{sylju@nordita.dk}

\thanks{}

\date{\today}

\preprint{NORDITA-2005-84}

\begin{abstract}
The plaquette phase of the square lattice quantum dimer model is studied using a continuous-time reptation quantum Monte Carlo method for lattices of sizes up to $48 \times 48$ sites. We determine the location of the phase transition between the columnar and plaquette phases to occur at $V_c/J=0.60 \pm 0.05$ which is significantly larger than inferred from previous exact diagonalization studies on smaller lattices. Offdiagonal correlation functions are obtained. They exhibit long-range order in the plaquette phase but not at the Rokhsar-Kivelson point. We also observe significant finite-size corrections to scaling for the transition between the plaquette phase and the critical resonating valence bond liquid. This study demonstrates the importance of understanding finite-size effects when considering critical properties of the square lattice quantum dimer model.
\end{abstract}

\maketitle

\section{Introduction}
The Quantum dimer models(QDMs) are interesting as they constitute simple examples of effective quantum lattice models with restricted Hilbert spaces. Quantum lattice models are frequently encountered in condensed-matter physics. They are defined by an Hamiltonian acting on a Hilbert space which is a direct product of Hilbert spaces for each site. 

In many cases the Hamiltonian consists of terms of widely different magnitude. The effects of the largest terms on the low-energy physics can then be effectively described by restricting the allowed Hilbert space. Sometimes this can be carried out at the level of the site Hilbert spaces, just restricting the allowed states on a single site. However in other cases the constraints cannot be modelled in this site-specific way and the constraint on the Hilbert space is more complicated. The QDMs are examples of the latter.

The QDMs were originally proposed\cite{KRS} as models for antiferromagnets where the tendency to form short-range spin-singlet valence bonds is strong. This formation of short-range valence bonds is modelled as a constraint which effectively reduces the Hilbert space to a nontrivial state space; that of dimer--coverings of the lattice\cite{well}. 

While the Hilbert space of a QDM is special its Hamiltonian is rather simple. It consists of a kinetic term that flips the orientation of parallel dimers and a potential energy that associates an energy cost/gain to parallel dimers. The QDMs can be formulated on any lattice and their phase diagrams are largely similar.  They consist of a liquid and various solid phases. The liquid phase is known as the resonating valence bond(RVB) liquid. The solid phases comes in at least three varieties, one phase with a maximum amount of parallel dimers; the columnar phase, one with no parallel dimers; the staggered phase, and one intermediate phase characterized by having a set of plaquettes with parallel dimers changing orientation constantly; a resonating plaquette phase. 

While there are good evidences for the existence of a columnar to a plaquette state phase transition in the QDM on the hexagonal\cite{Moessner01} and triangular\cite{Ralko05} lattices, the evidence is weaker on the square lattice where it is only based on exact diagonalization studies of linear system sizes up to $L=8$\cite{Leung}. In this article we show evidence of the columnar to plaquette phase transition for the square lattice QDM and estimate its location. We find that the plaquette phase is realized in a much smaller region in parameter space than previously estimated.

The resonating plaquette phase is most directly characterized by order in a quantity offdiagonal in the dimer basis that measures resonating dimers. We have therefore carried out simulations focusing on the possible appearance of long-range order in the offdiagonal dimer flip correlation function. While we find long-range-order in this quantity inside the plaquette phase, the magnitude of the order is significantly reduced from the value expected for an ideal plaquette product state with resonating dimers on one of the four plaquette sublattices, see Fig.~\ref{ideal}.

The ground state of a QDM is explicitly known at its Rokhsar-Kivelson(RK) point\cite{RK}. The particular form of the ground state implies that any T=0 quantum correlation function of observables that are diagonal in the dimer basis can be obtained as an infinite temperature correlation function of the classical dimer model. i.e they are just properties of the dimer-coverings themselves. This mapping has been utilized to calculate ground-state properties at the RK-point\cite{Fisher,MoessnerSondhi,Sandvik}. On the square lattice the ground state at the RK-point is critical. It was initially considered likely that the properties of the square lattice RK-point would extend also to the immediate vicinity of the RK-point\cite{RK}. However numerical diagonalization studies concluded that a significant portion of the phase diagram exhibits crystalline order\cite{Sachdev89} and that there were no evidence for a liquid state away from the RK-point\cite{Leung}. These studies were carried out on rather small lattices (up to $L=8$) and properties of the corresponding phase transition at the RK-point were not addressed. In this article we also attempt to address these properties by measuring the Binder ratio of the columnar order parameter close to the RK-point for system sizes up to $L=48$.  

The quantum Monte Carlo method employed in this article is a synthesis of the Continuous--time lattice Diffusion Monte Carlo method introduced in Ref.~\cite{Olav05} and the Reptation Monte Carlo method introduced in Ref.~\cite{BaroniMoroni}. This amalgam of methods which we term Continuous--Time Reptation Monte Carlo (CTRMC) is easy to implement and is free of population--bias and time--discretization errors that hamper various other forms of projector Monte Carlo techniques. The method is not restricted to QDMs and can be applied to any quantum lattice model free of the sign problem. 

Before we explain our results we will discuss the phase diagram of the square lattice QDM in greater detail.

\section{phase diagram}
The square lattice QDM Hamiltonian is
\be \label{Hamiltonian}
	H= -J \sum_{\rm plaq} \left(  \vphantom{\sum} | \verdimers \rangle \langle \hordimers | + \rm{H.c.} \right)
	   +V \sum_{\rm plaq} \left( \vphantom{\sum} | \verdimers \rangle \langle \verdimers | +
                                             | \hordimers \rangle \langle \hordimers |\right)
\ee
where the summations are taken over all elementary plaquettes of the lattice. We will choose units of energy such that the flipping energy $J=1$.

\begin{figure}
\mbox{
\includegraphics[clip,width=3cm]{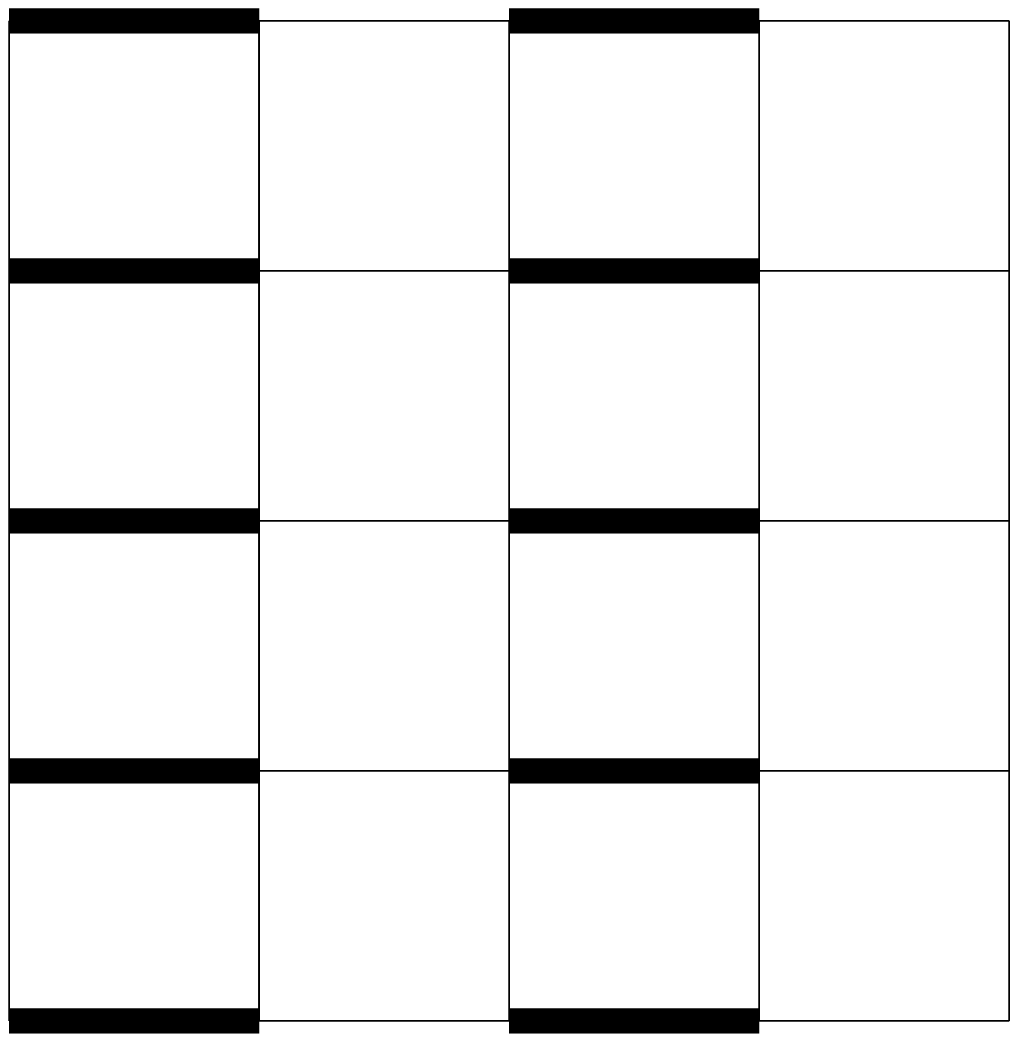}
\hspace{0.3cm}
\includegraphics[clip,width=3cm]{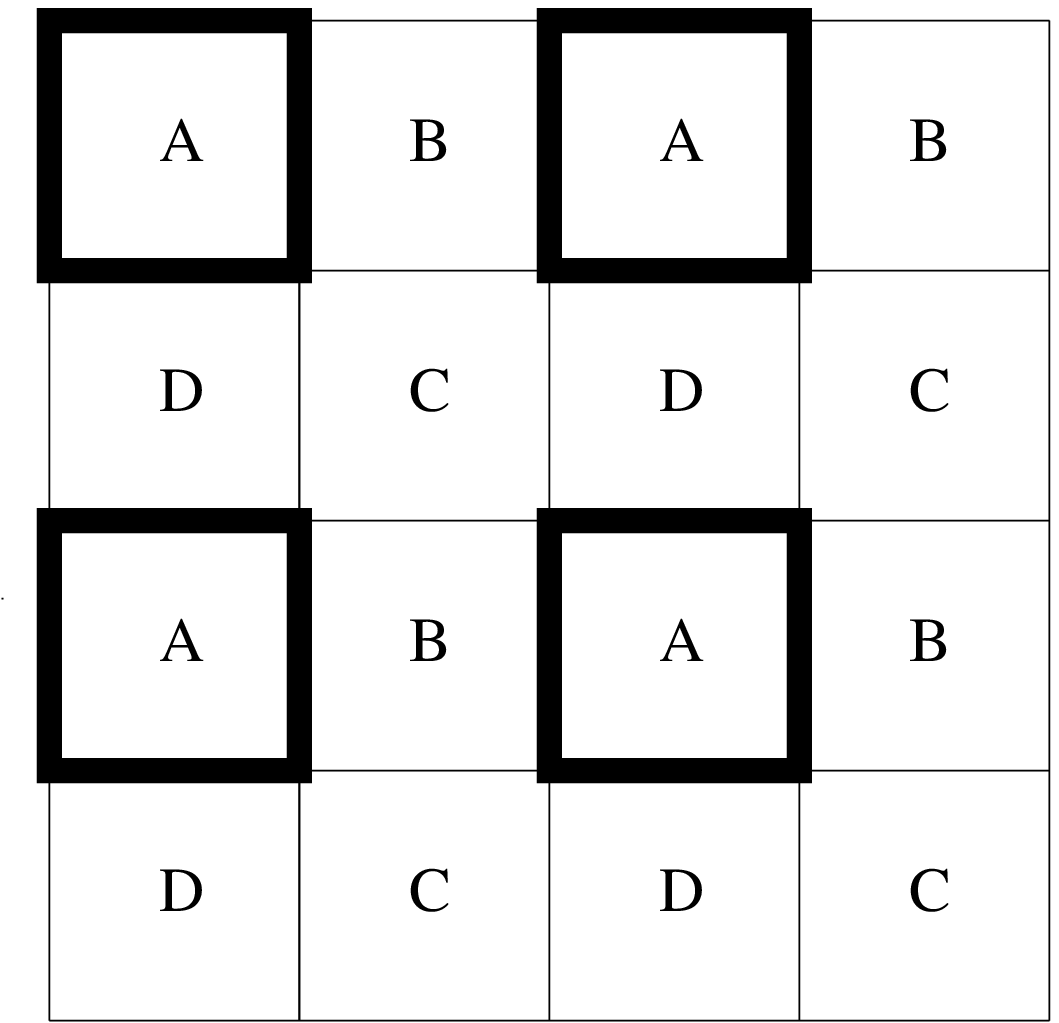}}
\caption{Schematic drawings of the ideal columnar state (left) and the ideal resonating plaquette state (right). The thick lines represent a high average dimer probability (1 for the ideal columnar state and 1/2 for the ideal plaquette state). The letters in the right panel show the assignment of the four different plaquette sublattices.
\label{ideal}}
\end{figure}

\begin{figure}
\includegraphics[clip,width=8cm]{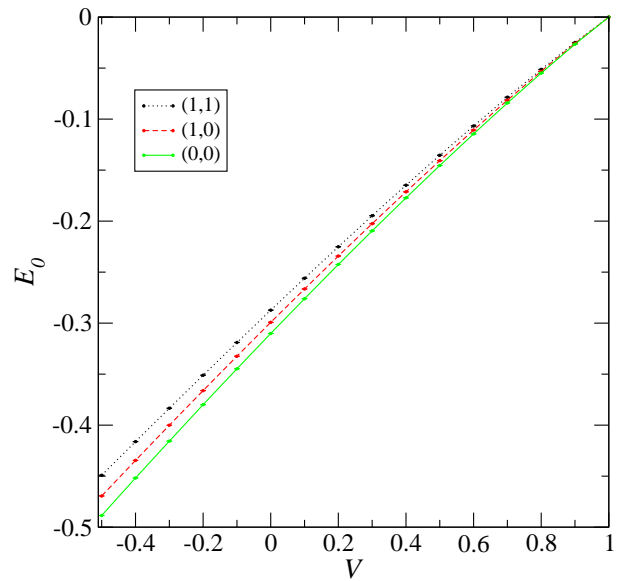}
\caption{Color online. Energy per plaquette for the lowest lying state in three different topological sectors characterized by the transition graph winding numbers $(w_x,w_y)$ with respect to the columnar reference configuration. From top to bottom the energy curves have winding numbers $(1,1)$,$(1,0)$ and $(0,0)$.
\label{erg}}
\end{figure}

The state space of the square lattice QDM is naturally divided into separate topological sectors each invariant under the action of the Hamiltonian. Any dimer configuration belongs to a topological sector characterized by the winding numbers of its transition graph to a reference configuration, which we take to be the ideal columnar state shown in Fig.~\ref{ideal} left panel. The transition graph is obtained by overlaying the reference configuration on the dimer configuration in question and erasing overlapping dimers. This leaves a set of loops which might wind around the lattice.  For $V<1$ the topological sector with zero winding numbers have the lowest energy, see Fig.~\ref{erg}, and we will restrict our simulations to this topological sector. 

\begin{figure}
\includegraphics[clip,width=6cm]{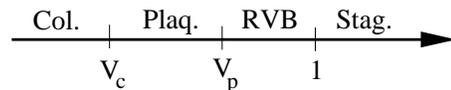}
\caption{Generic $T=0$ phase diagram of QDMs. The phases are from left: columnar phase, resonant plaquette phase, RVB liquid, staggered phase. For QDMs on bipartite lattices: $V_p=1$.   
\label{phasediagram}}
\end{figure}

A schematic zero temperature phase diagram of the QDMs is shown in Fig.~\ref{phasediagram}. For $V=1$, the RK-point, the ground state is the equal--amplitude superposition of all dimer coverings of the lattice. For the square lattice this implies that dimer-dimer correlations have no long-range order, but are critical and decay as a power law\cite{Fisher}. We will refer to this state as the RVB liquid although it is gapless and is believed to exist only at a single point in the phase diagram for QDMs on bipartite lattices\cite{Sachdev89,MoessnerSondhiFradkin}. For non-bipartite lattices like the triangular\cite{MoessnerSondhi,Ralko05} and the Kagome lattice\cite{Misguich02} the RVB liquid has gapped excitations and extends over a finite region in parameter space.

The QDMs exhibits a number of crystalline phases. For $V \to -\infty$ the ideal columnar state with a maximal number of parallel dimers will be preferred, see Fig.~\ref{ideal}. This four-fold degenerate (on a square lattice) state is diagonal in the dimer basis thus the kinetic term will tend to destroy it. These quantum fluctuations will for finite $V$ lead to ``disorder'' within the columns, but it is expected that the broken rotational symmetry of the ideal columnar state still survives at least up to a critical value of $V$. Thermal effects will also tend to destroy the $V\to -\infty$ columnar state, these were studied in Ref.~\cite{Alet}. It is reasonable to believe that the kinetic term will eventually, for big enough values of $V$, turn the ground state into a state resembling the ideal resonating plaquette state shown in Fig.~\ref{ideal}. The ideal resonating plaquette state is also four-fold degenerate on a square lattice and can be written as
 \be \label{ideal_plaq}
   | \psi_{\rm plaq} \rangle = \prod_{\rm A} \f{1}{\sqrt{2}} \left( | \verdimers \rangle + 
   | \hordimers \rangle \right)
\ee
where the product is taken over all plaquettes on one of the four plaquette sublattices. A similar resonant plaquette phase is known to exist for QDMs on the hexagonal\cite{Moessner01} and triangular\cite{Ralko05} lattices.

While the ideal columnar state is certainly the true ground state for $V \to -\infty$ there is no guarantee that the ideal plaquette state is the ground state for $V=0$ as the ideal plaquette state is not an eigenstate of all the kinetic terms. In fact it is only an eigenstate of the subset of kinetic terms acting on the plaquettes with resonating dimers. Nevertheless the $V=0$ point is believed to be situated inside the plaquette phase for both the hexagonal\cite{Moessner01}, triangular\cite{Ralko05} and the square\cite{Leung} lattices.

\begin{figure}
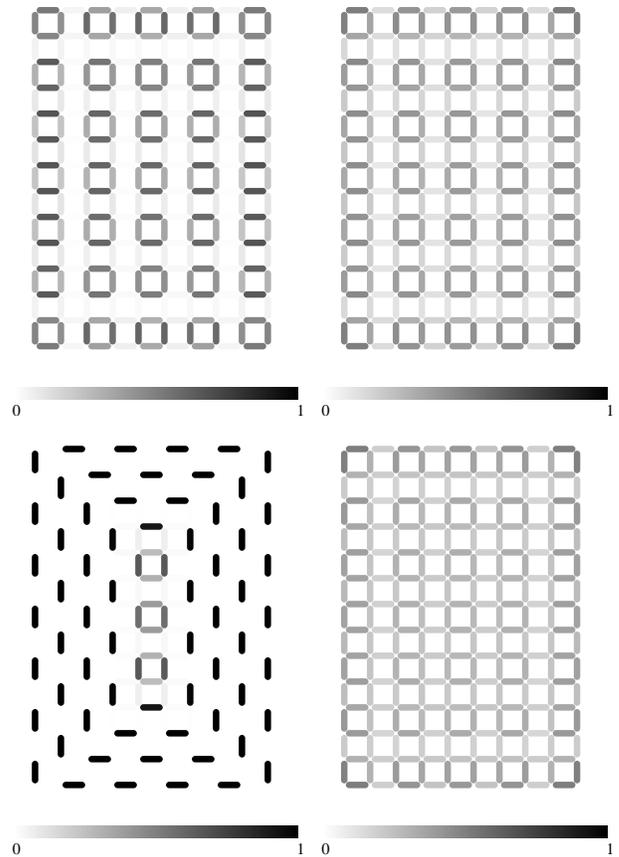

\mbox{
\includegraphics[clip,width=4cm]{fig4a}
\includegraphics[clip,width=4cm]{fig4b}
}
\mbox{
\includegraphics[clip,width=4cm]{fig4c}
\includegraphics[clip,width=4cm]{fig4d}
}
\caption{Average dimer densities on a rectangular lattice with open boundary conditions for different values of $V$. The gray-scale marks the dimer density. Clockwise from upper left: V=$-0.5$, $0.8$ ,$1.0$, $1.2$.
\label{avgdens}}
\end{figure}

It is clear that when $|V| < 1$ the ground state should have an appreciable amount of resonating dimers. In order to appreciate the deviations from the idealized states depicted in Fig.~\ref{ideal} and the significance of the resonating dimers we have plotted the average dimer densities for several values of $V$ in Fig.~\ref{avgdens}. These plots and Fig.~\ref{erg} were obtained using the Monte Carlo method which will be explained in the next section. In order to break translational and rotational symmetry the plots were obtained on a rectangular lattice with open boundary conditions. For $V<1$ (all panels except the bottom left in Fig.~\ref{avgdens}) one can see that a maximal number of flippable plaquettes is favored. This is seen as the darker squares in the plots. For $V=-0.5$ (upper left panel) the dimers on the central plaquette prefer to be horizontally aligned. We take this as an indication of the columnar phase although for the small system shown here it might as well be a boundary effect.  The orientational preference weakens as $V$ increases, and at $V=1$, the RK-point, the plaquette pattern is barely visible at all. In fact it disappears in the thermodynamic limit, and the remnant seen here is a finite-size effect. For $V>1$ a staggered arrangement of dimers is preferred. This is seen to be true locally in the lower left panel of Fig.~\ref{avgdens}. However the boundary conditions restrict the type of configurations that are available globally and the actual global pattern seen is a result of freezing into a specific configuration that depends on initial conditions and the exact pattern of quantum fluctuations. The $V>1$ phase of the square lattice QDM model is also very sensitive to perturbations as shown in Ref.\cite{Cantor}. We will consider $V \leq 1$ in the remainder of this article. 

The primary aim of this article is to find the location $V_c$ of the phase transition between the columnar and plaquette phases.  However first the quantum Monte Carlo method used in this article will be explained.

\section{Monte Carlo Method}
The object of CTRMC is to perform a stochastic simulation of the imaginary--time evolution operator $e^{-H\tau}$ which in the large time limit is a projection operator onto the ground-state of the Hamiltonian. As explained in details in Ref.~\cite{OlavIJMPB} it is possible to carry out this imaginary--time propagation entirely without time discretization errors for lattice systems.
 
To explain how the continuous--time propagation works in CTRMC consider first the possible actions of the evolution operator on a particular configuration during an infinitesimal time step $d\tau$. In CTRMC there are two possible actions: 1) The configuration changes or, 2) the configuration stays unchanged. In order to account for the in general non-Markovian nature of the imaginary--time evolution operator a weight is also associated to the evolving configuration. With this weight it is possible to enforce probability conservation for the two actions 1) and 2) provided the weight is altered if action 2) is chosen. This strategy is the same as utilized in so called pure (no branching) Diffusion Monte Carlo methods\cite{Hetherington,Caffarel}.

The probability of action 1) is given by offdiagonal elements of the Hamiltonian and is of the order $d\tau$. To void the sign problem these elements must all be negative or zero. The probability for the configuration to stay unchanged, action 2), is $p(2)=1-p(1)=1+\sum_i H_{ic}d\tau$, where $c$ labels the current configurational state and $i$ are other states connected to the current state by an offdiagonal term in the Hamiltonian. This choice of probabilities and the form of the infinitesimal time evolution operator, $1-H_{cc} d\tau$, implies that the multiplicative weight change associated with action 2) is $1- E_L(c) d\tau$ where $E_L(c)$ is the so called local energy, $E_L(c)=H_{cc} + \sum_i H_{ic}$, of configuration $c$.

Because $p(2)$ is of the order unity one can simulate the continuous--time evolution in the same way as done for continuous--time simulations of radioactive decay. One generates stochastically a future decay--time $\tau_{\rm decay}$ according to the distribution $e^{-(1-p(2))\tau_{\rm decay}/d\tau}$ and moves the configuration directly to this time while multiplying its weight with $e^{-E_L(c) \tau_{\rm decay}}$. At the decay time the type of decay, i.e. which state to transfer to, is determined dependent on the relative values of the many offdiagonal matrix elements leading away from configuration $c$. In this way the simulation is carried out without time-step errors. 

An evolving configuration will contribute to observables a term proportional to its accumulated weight. However it is well known that the weights will be widely spread in magnitude\cite{Assaraf00}. The sum will thus be dominated by just a few terms with the biggest weights thus giving essentially only the contributions for a few configurations yielding bad statistics. The advantage of the Reptation Monte Carlo technique is that this summation over configuration weights is {\em also} carried out stochastically using importance sampling based on the weight magnitudes. This sampling ensures a much more efficient summation as the observables gets relatively many contributions from the configurations with the highest weights.

\begin{figure}
\includegraphics[clip,width=8cm]{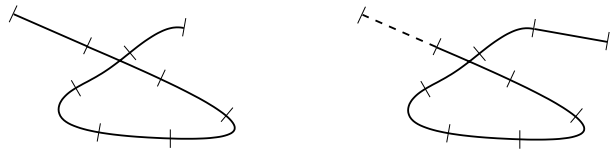}
\caption{Schematic drawing of two polymers or equivalently two propagations in state space. Each perpendicular line marks an element in the configuration array.  Each segment between two consecutive perpendicular lines indicates a $\Delta_\tau$ time propagation and has an associated entry in the weight array. The two polymers illustrate the evolution before(left) and after(right) the reptation move. The total weight of these two polymers differ only by their first and last segments. The dashed segment on the right polymer indicates a removed part so that both polymers have the same length.    
\label{newpolymer}}
\end{figure}

To make a practical implementation of CTRMC a starting configuration is stored as the first element in an array of configurations and then propagated a finite time interval $\Delta_\tau$. The resulting configuration is stored as the next element in the configuration array while the weight associated to the $\Delta_\tau$ propagation is stored as the first element of a separate array of weights. This process is repeated until the configuration has evolved for a total time $\tau_{\rm tot}= n \Delta_{\tau}$. The configuration array with $n+1$ elements and the weight array with $n$ elements constitute then a description of the continuous--time evolution. The total weight for this $\tau_{\rm tot}$ evolution is the product of weights in the weight array. The collection of the configuration and weight arrays will be referred to as the {\em polymer}. A schematic view of two polymers is shown in Fig.~\ref{newpolymer}.  
Having filled the polymer arrays the reptation move starts. First a random choice of propagation direction is made: Either the reptation propagation starts from the first element of the polymer and propagates backwards in time or from the last element propagating forward in time. Depending on the propagation direction the appropriate configuration (first or last) is copied and propagated for a time interval $\Delta_\tau$. After this step one essentially has information about two different evolutions that are almost identical except for their first and last elements. This is illustrated in Fig.~\ref{newpolymer}.

One now performs a Metropolis accept/reject decision based on the relative weights of the two polymers. That is: accept the reptation move with probability $p={\rm min}( W^\prime/W,1)$ where $W$($W^\prime$) is the total weight for the polymer before(after) the reptation step. This ratio of weights is easily calculated as it involves only the weights at the ends of the polymers. It depends clearly on the value of $\Delta_\tau$. We have found it most efficient to adjust $\Delta_\tau$ such that the resulting acceptance probability is about $1/2$.

Diagonal observables can easily be extracted from any stored configuration in the polymer. By extracting observables from the middle block in the polymer we ensure that observables are picked according to the forward-walking procedure\cite{forward} with a forward propagation time of $\tau_{\rm tot}/2$. 

The only adjustable parameter in this scheme is $\tau_{\rm tot}$ which ideally should be as long as possible. In this article we have taken it to be $\sim 40 J^{-1}$, and $80 J^{-1}$ close to the RK-point.

CTRMC can, as any other projection Monte Carlo techniques, be improved substantially by using a guiding function\cite{ImpSampling}. The guiding function should be as close as possible to the true ground state wave function. In our simulations we have chosen a guiding wave function that is biased towards having many plaquettes with parallel dimers. We optimize the guiding wave function by performing a small trial run before the actual run. In the trial run the guiding wave function is optimized in such a way as to yield a minimal variance of the local energy\cite{Umrigar88}. At the RK-point, where the ground state is explicitly known, the resulting quantum Monte Carlo simulation using the exact ground state as guiding wave function reduces to a classical Monte Carlo simulation which also can be employed to measure properties of exited states\cite{Henley97,Henley04,OlavIJMPB,Castelnovo}.

There is another class of Projector Monte Carlo methods where the non-Markovian character of the evolution operator is taken care of by introducing extra processes such as replication and decimation of copies of the system sometimes referred to as walkers\cite{Anderson,Trivedi}. Having such a changing number of walkers is undesirable as it will need some form of population control to prevent fluctuations from killing all walkers or filling the entire computer memory. Population control leads again to a systematic bias of the results which must be corrected for by reweighing\cite{Umrigar}. In CTRMC there is only one walker, thus these problems are avoided. It is also possible to get around these problems using a population with a stochastic reconfiguration of a constant  population of walkers as demonstrated in Ref.~\cite{Sorella}.

\section{Plaquette phase}
\subsection{Restoration of rotational symmetry}
The columnar ordering phase is in part distinguished from the plaquette phase by having a preferred dimer orientation. Thus a suitable order parameter distinguishing these phases is one that detects any orientational preferences of the dimers. Such an order parameter can be constructed by finding an operator that changes sign under a $\pi/2$ lattice rotation and a subsequent translation. The translation is needed to make the plaquette state invariant under this transformation.  

Leung {\em et al.}\cite{Leung} proposed to detect this by measuring the difference between the number of vertical and horizontal dimers
\be
	M_{\rm vh} = \f{2}{L^2} \left( N_{\rm v} -N_{\rm h} \right).
\ee
Here $N_{\rm v}$($N_{\rm hh})$ is the total number of vertical (horizontal) dimers. In the ideal columnar state $M_{\rm vh}^2 =1$ whereas in the ideal plaquette state $M_{\rm vh}^2 =4/L^2$ which vanishes in the thermodynamic limit. $M_{\rm vh}$ is therefore suitable as an order parameter.
 
Leung {\em et al.} studied $M_{\rm vh}$ and its Binder ratio up to system sizes $L=8$ and concluded that there is a  phase transition at $V_c=-0.2$. However the crossing points of the Binder ratios vary significantly with system size so bigger systems are needed to determine the location of the phase transition more accurately. Progress on bigger system sizes, up to $L=20$  were obtained in Ref.~\cite{OlavIJMPB} using a Diffusion Monte Carlo method, but no finite-size analysis of the data leading to a definite estimate for the location of the phase transition were given. 

\begin{figure}
\includegraphics[clip,width=8cm]{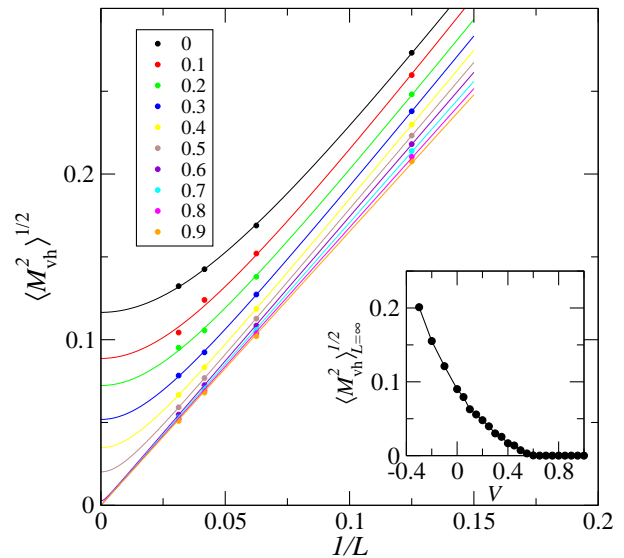}
\caption{Color online.  The order parameter $\langle M_{\rm vh}^2 \rangle^{1/2}$ vs. $1/L$ for different values of $V$. The top curve has $V=0$ and increases in steps of $0.1$ for lower curves. 
The solid lines are fits to the form $\sqrt{ \mu + \alpha/L^2}$. The inset shows the values of $\mu$ vs. $V$. The inset contains data for more values of $V$ than shown in the main figure. 
\label{nvh}}
\end{figure}

In order to find the location of the phase transition we have extracted $\langle M_{\rm vh}^2 \rangle^{1/2}$ from our simulations for different system sizes up to $L=32$. In Fig.~\ref{nvh} we have plotted the results for $\sqrt{\langle M_{\rm vh} \rangle^2}$ as a function of $L$ for different values of $V$. The solid lines are best fits to the functional form $\sqrt{\mu + \alpha/L^2}$. In the inset we have plotted the resulting infinite size extrapolation value ($\mu$)  where also results from other values of $V$ is included. From this figure we get our best estimate for the location of the phase transition between the columnar and plaquette state: $V_c = 0.60 \pm 0.05$. 

\begin{figure}
\includegraphics[clip,width=8cm]{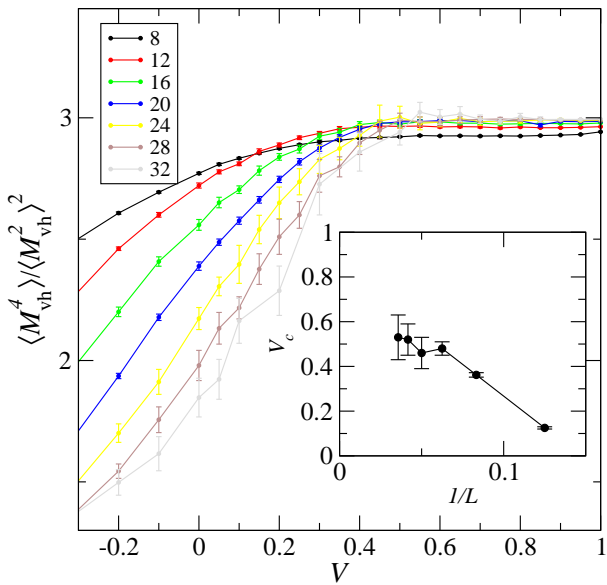}
\caption{Color online. Binder ratios vs. $V$ for different system sizes. The
top curve on the left side corresponds to linear system size $L=8$. $L$ increases successively by 4 for lower curves (on the left site). The inset marks the crossing points for $L$ and $L+4$ as functions of $L$. 
\label{Binderratio}}
\end{figure}

We have also extracted higher moments of the $M_{\rm vh}$ distribution. In Fig.~\ref{Binderratio} we show the Binder ratio $\langle M_{\rm vh}^4 \rangle/\langle M_{\rm vh}^2 \rangle^2$ as functions of $V$ for different system sizes. For large values of $V$ all curves have values close to 3, consistent with a Gaussian distribution with zero mean. The curves for different system sizes do not cross in a single point. However, as seen in the inset of Fig.~\ref{Binderratio}, the crossing points for curves of system sizes $L$ and $L+4$ seems to converge to a point close to $V_c=0.6$ consistent with the estimate above. However the rather large error bars on the crossing points do not constrain this estimate further. 

The phase transition between the columnar and plaquette phases on the square lattice is often conjectured to be a first order transition. This is consistent with what happens on the hexagonal lattice\cite{Moessner01} and in other models with similar phase transitions\cite{Shannon,OlavSudip}. The limited system sizes studied here makes it however rather difficult to verify this conjecture. We do not observe any hints of discontinuities in the differentiated ground state energy for the largest system sizes. Neither do we see any discontinuities in the way the order parameter approaches zero. However this does not rule out a weak first order transition that might only be visible in simulations using larger lattices.

\subsection{Breaking of translational symmetry}
The ideal columnar state in Fig.~\ref{ideal} is invariant under translation by one lattice spacing perpendicular to the orientation direction of the dimers. The plaquette phase breaks this symmetry. One could imagine an intermediate phase that breaks both this translational symmetry as well as the rotational symmetry discussed in the previous section. A phase like this could resemble that shown in the upper left panel of Fig.~\ref{avgdens} and be described for instance as the product state in Eq.~\ref{ideal_plaq} but with different amplitudes for the vertical and horizontal directions. 

\begin{figure}
\includegraphics[clip,width=8cm]{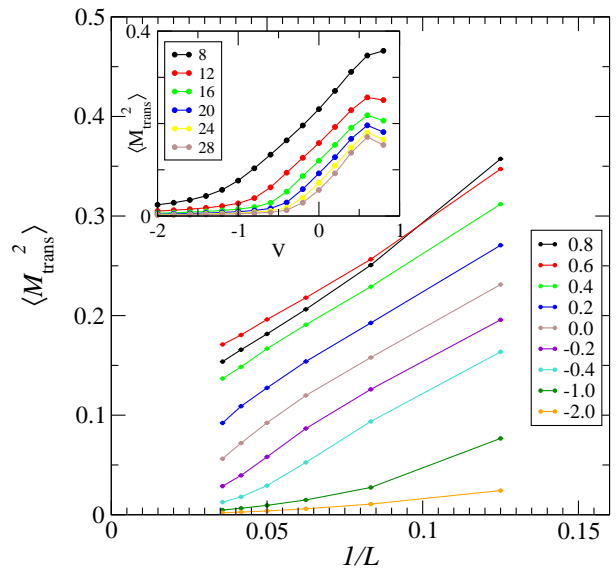}
\caption{Color online. Main panel shows the translational symmetry breaking order parameter vs. inverse linear system size for different values of $V$. The values are for the curves from top to bottom on the right side $V=0.8,0.6,0.4,0.2,0.0,-0.2,-0.4,-1.0,-2.0$.
 The inset shows the same order parameter, but plotted as a function of $V$ for different system sizes. Here the smallest system size is for the top-most curve.  \label{transsize}}
\end{figure}

An order parameter which detects the breaking of translational symmetry by one lattice spacing in the direction perpendicular to the orientation of the majority of dimers is
\bea
   M_{\rm trans} & = & \f{8}{L^2} \sum_{\rm plaq} \left[  \theta(M_{\rm vh}) N_{\rm vv}(\vec{r})(-1)^{r_x} \right. \nonumber \\
   &  &+ \left. \theta(-M_{\rm vh}) N_{\rm hh}(\vec{r})(-1)^{r_y} \right]
\eea
where $N_{\rm hh}$ ($N_{\rm vv}$) takes the value one for a plaquette with two horizontal (vertical) dimers and zero otherwise. $\theta(M_{\rm vh})=1$ if there are more vertical than horizontal dimers and 0 otherwise, $\theta(-x)=1-\theta(x)$. $\vec{r}$ denotes the midpoint coordinate of a plaquette; its component are integers in units of the lattice spacing. $M_{\rm trans}$ is invariant under $\pi/2$ lattice rotations and changes sign when a configuration is translated one lattice spacing along the direction perpendicular to the orientation of the majority of dimers. Thus $M_{\rm trans}=0$ in the columnar phase, while it takes the value $\pm 1/2$ for the ideal plaquette states. The inset of Fig.~\ref{transsize} shows $\langle M_{\rm trans}^2 \rangle$ vs. $V$ for different system sizes. For a fixed system size this order parameter stays small for low $V$ and increases towards a maximum at about $V=0.6$ and then decreases again towards $V=1$. The decrease at high values of $V$ is partly due to the reduced number of flippable plaquettes. For all the system sizes investigated the order parameter decreases with system size. 

In order to see the finite-size behavior of this decrease we have plotted the order parameter as a function of inverse linear system size in the main panel of Fig.~\ref{transsize}. It appears that the order parameter extrapolates to 0 for $V \leq 0.2$. Based on this we rule out translational symmetry breaking for $V \leq 0.2$. A close inspection of the $V=0.4$ curve reveals a downward curvature similar to that seen more clearly for the $V=0.2$ curve, thus we believe that the $V=0.4$ will also extrapolate to zero. In contrast the $V=0.6$ curve reveals an upward curvature indicating an extrapolation to a finite value and therefore breaking translational symmetry. The $V=0.8$ curve lies lower than the $V=0.6$ but has also a slight upward curvature. Thus we conclude that the translational symmetry breaking happens for $V\sim 0.4-0.6$. However larger system sizes are necessary to determine a more precise value. 

A critical value of $V\sim 0.4-0.6$ is close to the value of $V$ where rotational symmetry gets restored as found in the previous section.
Thus if a phase that breaks both translational and rotational symmetry exists, it can only do so in a rather narrow region of $V$ close to the rotational symmetry breaking transition.  

\section{Offdiagonal correlation functions}
\begin{figure}
\includegraphics[clip,width=8cm]{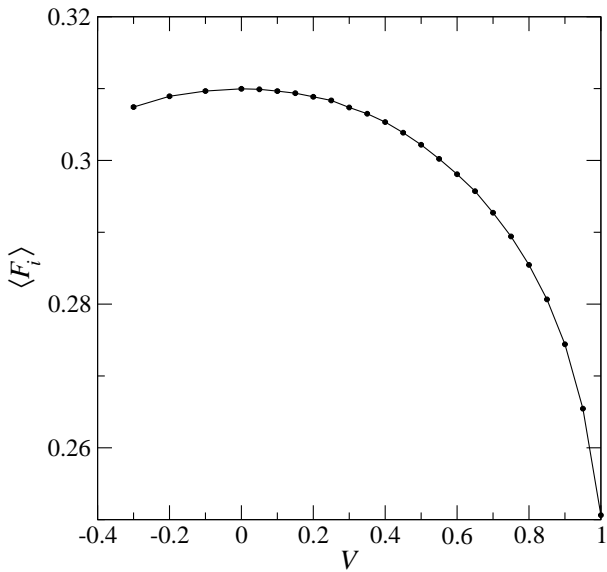}
\caption{Average value of flip operator for different values of $V$.  
\label{flip}}
\end{figure}

The resonating plaquette phase is most directly probed by looking for resonating dimers on one of the four sublattices. A resonating plaquette is characterized by a finite expectation value for the off-diagonal flip operator 
\be
	F_i = | \hordimers \rangle_i \langle \verdimers | + | \verdimers \rangle_i \langle \hordimers |, 
\ee
where $i$ labels the plaquette. 
To get an impression of how much the dimers are resonating we plot in Fig.~\ref{flip} the average value of the flip operator per plaquette. This can be calculated
directly from measuring the ground state energy per plaquette $E$ and the potential energy
\be
	\f{1}{L^2} \sum_i \langle F_i \rangle = -E + V \langle \f{N_f}{L^2} \rangle,
\ee
$N_f$ is the number of plaquettes with parallel dimers in a given configuration. We see that the dimers resonate the most at $V=0$.

To get a picture of how the resonating dimer plaquettes are correlated on the lattice we measure the correlation function $\langle F_i F_j \rangle$. In the plaquette phase this correlation function should show long range order when $i$ and $j$ both belong to the sublattice with resonating dimers. In order to measure $\langle F_i F_j \rangle$ we used the Feynman-Hellman theorem. An operator $ -\alpha F_i F_j$ was added to the Hamiltonian and the ground state energy was measured as a function of $\alpha$. Taking small values of $\alpha$ the expectation value of $F_i F_j$ was determined as $\langle F_i F_j \rangle = -\f{\partial E_g}{\partial \alpha} |_{\alpha \to 0}$.

\begin{figure}
\includegraphics[clip,width=8cm]{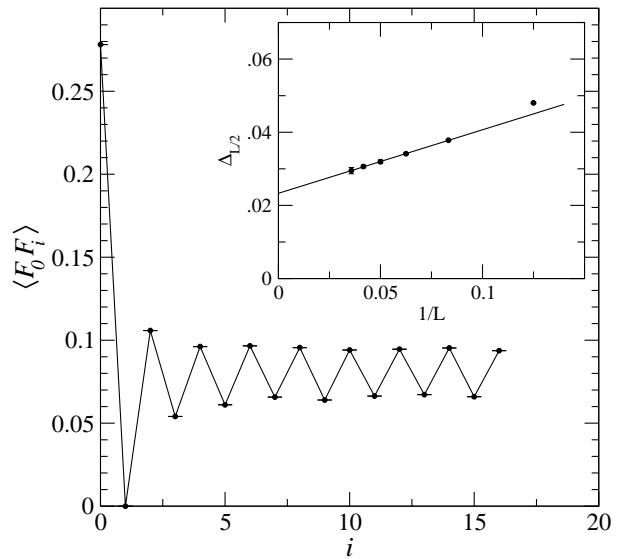}
\caption{Offdiagonal correlation function $\langle F_0 F_i \rangle$ as function of diagonal separation $\vec{r}_i = (i,i)$ at $V=0.9$ on a $L=32$ lattice.
The inset shows the behavior of $\Delta_{L/2}$ for different values of $L$. The line in the inset is the best linear fit to the largest system sizes (excluding $L=8$) and is $\Delta_{L/2} = 0.023 + 0.17/L$.
\label{offd09}}
\end{figure}

In Fig.~\ref{offd09} we have plotted $\langle F_0 F_i \rangle$ for $V=0.9$ at
diagonal spatial separations: $i$ denotes a plaquette with coordinates $(i,i)$.
The zero result for $i=1$ is an exact result as two diagonally adjacent plaquettes cannot both be flippable. The curve exhibits clear oscillations in the bulk indicative of a resonating plaquette phase. To find out if these oscillations are also present in the thermodynamic limit, we have defined a quantity $\Delta_{L/2}$ that measures the magnitude of the oscillations in the bulk: $\Delta_{L/2} = \langle F_0 F_{L/2}\rangle - \langle F_0 F_{L/2-1} \rangle$. In the inset of Fig.~\ref{offd09} we have plotted $\Delta_{L/2}$ as a function of $1/L$. $\Delta_{L/2}$ extrapolates to a finite value in the thermodynamic limit, consistent with the presence of a resonating plaquette phase. One should however note that the magnitude of these bulk oscillations are small compared to what is expected in the ideal plaquette state, Eq.~\ref{ideal_plaq}. A $\Delta_{\infty}=0.023$ for $V=0.9$ corresponds to $9\%$ of the value expected for one of the ideal plaquette states.

\begin{figure}
\includegraphics[clip,width=8cm]{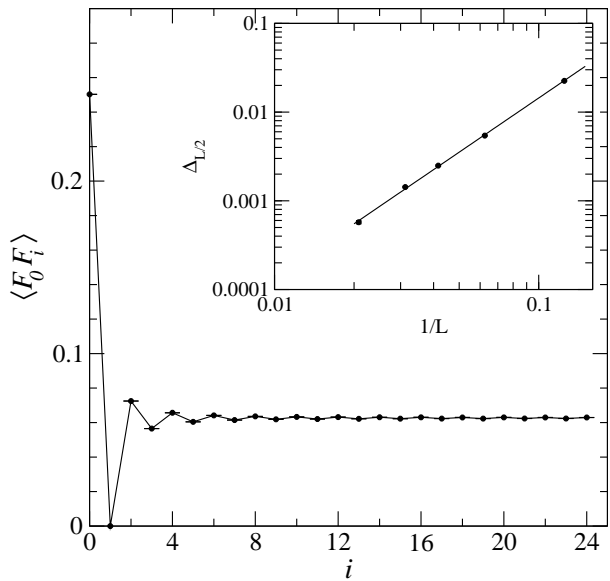}
\caption{Offdiagonal correlation function $\langle F_0 F_i \rangle$ as function of diagonal separation $\vec{r}_i = (i,i)$ at the RK-point on an $L=48$ lattice.The inset shows $\Delta_{L/2}$ vs. $1/L$. The line is the best fit to a power law $\Delta_{L/2} \sim 1/L^g$ with $g=2.02 \pm 0.03$.
\label{offdRK}}
\end{figure}

To contrast this finding at $V=0.9$ we have repeated the same measurements at the RK-point. The obtained result for $\langle F_0 F_i \rangle$ at the RK-point is shown in Fig.~\ref{offdRK} for an $L=48$ lattice. Although visible the bulk oscillations are much smaller in this case. In fact they vanish completely in the thermodynamic limit as can be seen from the inset of Fig.~\ref{offdRK} which shows the finite-size scaling of $\Delta_{L/2}$ which vanishes as a power law $L^{-g}$. The bulk decay at the RK-point can be calculated analytically  by calculating the correlation function for having two parallel dimers on a plaquette displaced from another plaquette also with two parallel dimers. Using the Pfaffian technique\cite{Fisher} and the Green function given in Ref.~\cite{Fendley} it is easy to show that asymptotically $\langle F_0 F_i \rangle \sim (-1)^i/i^2$, thus the exact value of $g=2$. Our numerical result is consistent with this.      

\section{Transition to RVB liquid}
The columnar order parameter
\bea
	M^2_{\rm col}  & = & \f{1}{4L^4} \left[ 
	\left(\sum_{\rm plaq} N_{\rm h}(\vec{r}) (-1)^{r_x}\right)^2  \right. \nonumber \\
	& +& \left. \left(\sum_{\rm plaq} N_{\rm v}(\vec{r}) (-1)^{r_y}\right)^2 \right], 
\eea
where $N_{\rm h}(\vec{r})$ ($N_{\rm v}(\vec{r})$) is the number of horizontal (vertical) dimers surrounding the plaquette at $\vec{r}$,  was proposed in Ref.~\cite{Sachdev89} as a mean to detect the columnar order. $|M_{\rm col}|=1/2$ in the ideal columnar state. The columnar order parameter is constructed so that it is $0$ for any state invariant under lattice rotations. The plaquette state is not invariant under lattice rotations alone. It is invariant under the combined operation of a rotation and a translation, thus in fact the columnar order parameter for the ideal plaquette state is $|M_{\rm col}|=1/\sqrt{8}$ in the thermodynamic limit. Therefore the columnar order parameter does not distinguish between the columnar and plaquette phases in a useful way. However the columnar order parameter is zero in the RVB liquid and in the staggered state, thus the phase transition from the plaquette phase to the RVB liquid can be detected using the columnar order parameter.

\begin{figure}
\includegraphics[clip,width=8cm]{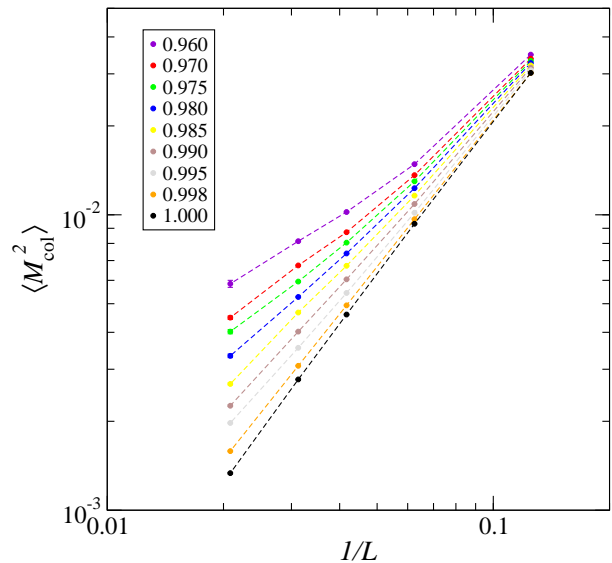}
\caption{Color online. $\langle M_{\rm col}^2 \rangle$ vs. $1/L$ for various values of $V$. From top to bottom the values of $V$ are $V=0.960,0.970,0.975,0.980,0.985,0.990,0.995,0.998,1.000$. The dashed lines are guides to the eye.
\label{colRK}}
\end{figure}

In Fig.~\ref{colRK} we show the size dependence of $M_{\rm col}^2$ for different values of $V$ close to the RK-point. For all curves the order parameter decreases with increasing system size. As $V$ is moved away from $1$ the decrease becomes slower and appears to saturate to a finite value, at least for the curves with $V<0.99$. This saturation to a finite value is not manifest for the curves with $V$ close to $1$ for the system sizes considered here. 

\begin{figure}
\includegraphics[clip,width=8cm]{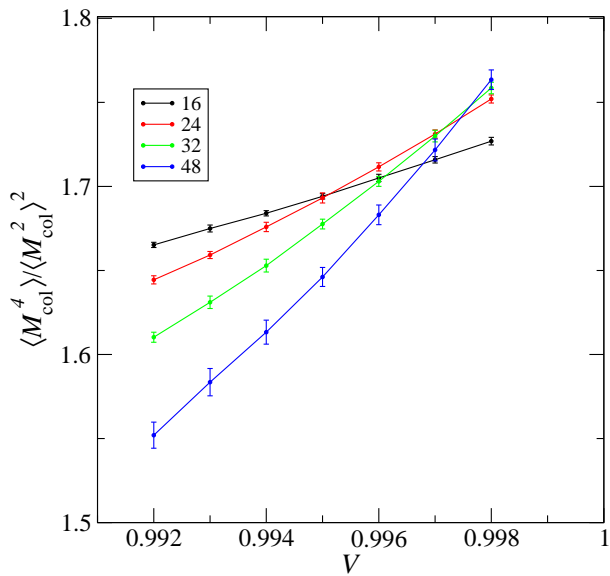}
\caption{Color online. The Binder ratio $\langle M_{\rm col}^4 \rangle/ \langle M_{\rm col}^2 \rangle^2$ vs. $V$ for different system sizes. From top to bottom on the left side the curves are for sizes $L=16,24,32$ and $48$.
\label{colBinder}}
\end{figure}

In order to search for a possible phase transition at $V < 1$ we show the Binder ratio $\langle M_{\rm col}^4 \rangle/\langle M_{\rm col}^2 \rangle^2$ as functions of $V$ for different system sizes in Fig.~\ref{colBinder}. The curves for different sizes do not cross in a {\em single} point. There is a tendency that the crossing points of curves for nearby system sizes move towards 1 as the system size increases. Interpreted this way we conclude that there is no evidence for a phase transition to the RVB liquid at $V < 1$ and that there is significant finite-size corrections to scaling even for the largest system sizes considered here.   

\begin{figure}
\includegraphics[clip,width=8cm]{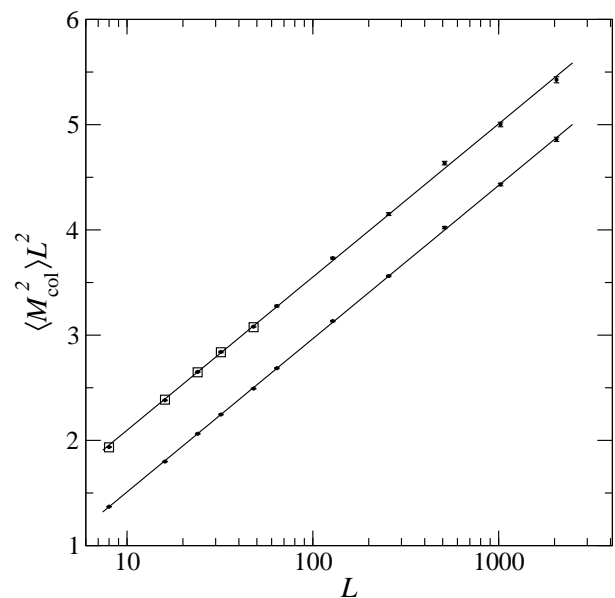}
\caption{Directed-loop Monte Carlo simulation results showing $\langle M_{\rm col}^2 \rangle L^2$ vs. $L$ at the RK-point on a semi-log plot. In the upper curve the data points are obtained by restricting the sampling to the zero winding number sector. The squares are centered about the Reptation Monte Carlo results gotten from Fig.~\ref{colRK}. In the lower curve the data points are collected from all topological sectors. The lines are best fits to the data using the functional form $\langle M_{\rm col}^2 \rangle = C \log(L)/L^2$ and gives within error bars the same $C=0.63 \pm 0.01$ for both curves. The largest system size simulated had $L=2048$.
\label{loop}}
\end{figure}

To check the columnar order parameter results at the RK-point we have in addition employed a very effective directed-loop Monte Carlo method\cite{Sandvik} which only is applicable exactly at the RK-point. Fig.~\ref{loop} shows the results plotted so as to expose the logarithm present in the leading asymptotically result $\langle M_{\rm col}^2 \rangle = C \log(L)/L^2$. This result can be calculated analytically using the Pfaffian technique with the Green function given in Ref.~\cite{Fendley}. The directed-loop Monte Carlo results agree with this asymptotic behavior, see Fig.~\ref{loop}, both when all topological sectors are included in the sampling (lower curve) and when the sampling is restricted to the zero winding number sector (upper curve). We find that the value of $C = 0.63 \pm 0.01$. Fig.~\ref{loop} also shows the CTRMC data from Fig.~\ref{colRK} (squares). They coincide with the directed-loop Monte Carlo data restricted to the zero winding number topological sector.

\section{Conclusion}
We have applied a continuous-time variant of the reptation quantum Monte Carlo method to study the square lattice QDM. In particular we found the location of the phase transition between the columnar and plaquette phase to be at $V_c = 0.60 \pm 0.05$. This phase transition happens at a positive $V_c$ and excludes the resonating plaquette state as the ground state when the Hamiltonian consists of just the kinetic term. 

The estimate for the location of the phase transition is based on the finite-size extrapolation of the order parameter and the apparent convergence of the Binder ratios. Thus there is a possibility that our estimate is strictly a lower bound as we cannot rule out a scenario were the order parameter is very small but finite, and where a phase transition is only seen for very large lattices.  Our results shine little light on the thermodynamic properties of the phase transition. The crossing points for the Binder ratio of the rotational symmetry breaking order parameter converge rather slowly, thus there are significant finite-size corrections to scaling if the transition is continuous. If the transition is first order, it is rather weakly so, as we see no evidences for discontinuities in the differentiated ground state energy nor in the order parameter for the system sizes considered here.

That finite-size effects are very important in these studies is also apparent from the average dimer densities in open boundary geometries shown in Fig.~\ref{avgdens} where one can clearly see plaquette patterns both far into the columnar phase and at the RK-point. 

We have also considered the possibility of an intermediate phase in between the columnar and plaquette phase breaking both rotational and translational symmetries. In order to do so we have constructed an order parameter that is insensitive to rotations, but detects the breaking of translational symmetry in the direction perpendicular to the orientation of the majority of dimers. We find a phase transition occurring at roughly the same value of $V$ as the rotational symmetry gets restored. Thus if such an intermediate phase exists it is confined to a narrow region in phase space close to $V \sim 0.6$.

To strengthen our argument for the existence of the plaquette phase we have also measured the off-diagonal dimer-flip correlation function inside the plaquette phase and performed a finite-size scaling of the results. This reveals long-range (staggered) order in the thermodynamic limit consistent with the existence of a plaquette phase with resonating dimers. However the strength of the plaquette pattern seen at $V=0.9$ is rather weak. It is only about $9\%$ of the ideal plaquette state value. At the RK-point we find that the plaquette pattern vanishes as the square of the linear system size in agreement with analytic calculations.  

We have also considered the transition from the plaquette phase to the RVB liquid as measured by the columnar order parameter.
To verify the CTRMC data for the columnar order parameter at the RK-point we measured the columnar order parameter using a directed-loop Monte Carlo algorithm. These methods gave the same results, and the directed-loop Monte Carlo algorithm was employed to verify the analytic prediction for the finite-size scaling of the columnar order parameter up to linear system sizes $L=2048$.   

From the columnar order parameter data away from the RK-point, there are no evidences for a phase transition occurring for $V<1$. The crossing points of the Binder ratios for different system sizes from $L=16$ to $L=48$ move towards higher values of $V$ as the system sizes are increased. Thus we conclude that there is significant finite-size corrections even at the largest system sizes considered here. Large finite-size effects is to be expected as the effective height-model describing the system close to the RK-point contains a dangerous irrelevant operator\cite{Henley97,Cantor}, making the extraction of scaling parameters from finite sized samples a complicated issue\cite{Luijten}. More numerical work on larger lattices combined with a proper finite-size scaling ansatz is needed to extract the proper critical behavior of this phase transition.

\begin{acknowledgments}
The author acknowledges helpful discussions with Anders Sandvik. Part of this work was supported by the Boston University visitors program. The computer simulations were  in part carried out using computers made available by the Department of Computer Science at Aalborg University thru the Nordugrid project.    
\end{acknowledgments}

 \end{document}